\documentclass{ws-procs961x669}   
\usepackage{slashed}
\usepackage{mathtools, amssymb}
\begin{document}
\title{Anomalies, CPT and Leptogenesis\\
}

\author{Sarben Sarkar$^*$ }

\address{Theoretical Particle Physics and Cosmology Group,\\ Department of Physics, King's College London,\\
London, WC2R 2LS, UK\\
$^*$E-mail: sarben.sarkar@kcl.ac.uk\\
www.kcl.ac.uk}



\begin{abstract}
We bring together different puzzles in physics beyond the Standard Model of particle physics. Within our model   baryogenesis, neutrino mass, strong CP and dark matter puzzles are related. The common ingredient in connecting these puzzles is the Kalb-Ramond field, a two form present in the gravitational multiplet in the theory of closed strings. Leptons are fermions which we need to couple to gravitational degrees of freedom using a vierbein formalism. The presence of torsion provided by a Kalb-Ramond background leads us to firstly an effective model involving the Einstein-Cartan formulation of gravity, a gauge theory, in which coupling to fermions is via a covariant derivative and secondly to a mechanism for CPT violation. This picture emerges from a low energy string effective action obtained from a closed bosonic string theory after compactification to four dimensions. The Kalb-Ramond field in four dimensions can be identified with a pseudoscalar gravitational axion. Because of the presence of an axial anomaly this axion can couple with the gluon field, and in this way allows a connection with the strong CP problem and axionic dark matter.

\end{abstract}

\keywords{Axion, Kalb-Ramond field, Torsion, Leptogenesis}

\bodymatter

\section{Baryogengesis}\label{aba:sec1}
Our main puzzle is baryogenesis. The observed baryon asymmetry (i.e. there is hardly any primordial antimatter) in the current era implies that, for an earlier era (for times $t<10^{-6}$s),
\begin{equation}
\label{e1}
\Delta n_{B}\left(T\sim1GeV\right)=\frac{n_{B}-n_{\overline{B}}}{n_{B}+n_{\overline{B}}}\sim(8.4-8.9)\times10^{-11}
\end{equation}
where $n_B$ is the baryon number density and $n_{\overline{B}}$ is the anti-baryon number density at an early epoch of the expansion of the universe . An asymmetry $\Delta n_{L}$ between leptons and anti-leptons is also expected to be of the same order of magnitude as $\Delta n_{B}$. The requirement of anomaly cancellation  correlates $\Delta n_{B}$ and $\Delta n_{L}$.

 Within the Standard Model (SM) it is \emph{not} possible to find this level of baryon asymmetry- a quantitative issue. Indeed, assuming CPT~\cite{R1a} invariant theories, Sakharov~\cite{R1}  suggested that the following criteria would  need to be met for a theory to allow for baryogenesis:
\begin{enumerate}
  \item The existence of non-perturbative baryon number violating processes in the SM
  \item C and CP violation~\cite{CP} should be present so that the amplitudes for the processes 
  $X\rightarrow Y+b$ and $\bar{X}\rightarrow\bar{Y}+\bar{b}$ are unequal.
   \item Out of equilibrium processes
\end{enumerate}
The observed $CP$ violation in the neutral kaon experiments is insufficient to obtain the result of (\ref{e1}). Although ideas for increasing the amount of CP violation abound~\cite{R2} there is no accepted consensus on a favoured idea since there is fine tuning; so the framework of Sakharov remains influential but qualitative.
\
\subsection{CPT violation}

CPT invariance~\cite{R1a} is a mainstay of  local Lorentz invariant theories in flat space-time. The CPT theorem states that, for a Hermitian local Lorentz invariant theory, CPT is a symmetry where the discrete symmetries $C$, $P$ and $T$ are charge conjugation, parity and time-reversal~\cite{schwartz} for a system with scalar $\phi$, spinor $\psi$ and vector fields $A_{\mu}$. A particularly straightforward proof has been given by Luders\cite{R1b}. However, in the presence of gravity since we only have local Lorentz invariance, there is no reason for the theorem to be necessarily true and Lorentz invariance to hold. This is compounded if gravity can be quantised in some theory at the Planck scale. Space and time are likely to be granular.

A phenomenological renormalisable quantum field theory, which is an extension of SM and takes into account violation of Lorentz invariance (LIV) and the CPT theorem (CPTV), is known as the Standard Model extension~\cite{R15} (SME). The free fermion Lagrangian $L_f$ becomes
\begin{eqnarray}
L_{f} & = & i\bar{\psi } \left( \gamma_{\mu } +c_{\mu \nu }\gamma^{\nu } +d_{\mu \nu }\gamma_{5} \gamma^{\nu } +e_{\mu }+if_{\mu }\gamma_{5} +\frac{1}{2} g_{\mu \nu \rho }\sigma^{\nu \rho } \right)  \partial^{\mu } \psi  \nonumber\\
&  &-\bar{\psi } \left( m+a_{\nu }\gamma^{\nu } +b_{\nu }\gamma_{5} \gamma^{\nu } +\frac{1}{2} H_{\nu \rho }\sigma^{\nu \rho } \right)  \psi 
\end{eqnarray}
where $a_{\mu },b_{\mu },c_{\mu \nu },d_{\mu \nu },e_{\mu },f_{\mu },g_{\mu \nu \rho }$ and $H_{\mu \nu }$ are indexed constants which do not transform under Lorentz transformations.
String theory~\cite{R17}, which has promise of providing a quantum theory of gravity, will lead us to a related model which will no longer be purely phenomenological. The term of interest to  us is $b_{\mu }$which will become a quantum axion field and will also have a time-dependent background~\cite{R9, R14, R21}. In the proof of the CPT theorem it is assumed that the coupling constants are not functions of space-time. Such space-time dependent couplings can arise if there are non-trivial  backgrounds of gravitational or other tensorial fields and so may lead to CPTV.The formation of such backgrounds is thus interesting.
CPT is a symmetry between particles and antiparticles and so its breaking is an  issue for the understanding of baryogenesis.

 We will now briefly first summarise the nature of the puzzles to which we can relate baryogenesis or its related cousin leptogenesis.
 
 \subsection{Neutrino mass}
 
 Although the electron neutrino was discovered in the 1950s, the details of its extended family structure is still providing challenges to the SM~\cite{SM}.  There is evidence for oscillations between different species of neutrinos~\cite{Neutrinos}, which requires them to have tiny masses. The SM matter content was chosen to ensure that neutrinos remained massless to all orders in perturbation theory. Consequently attempts have been made by extending the field content of SM through the inclusion of right-handed Majorana sterile neutrinos~\cite{Fukugita} for example.  With the help of heavy sterile neutrinos, it is possible in tree-level analysis, to generate sufficiently small neutrino masses which may be compatible with neutrino oscillation data. This is known as the see-saw mechanism~\cite{Mohapatra}. However higher order effects (in the absence of a new symmetry) are likely to make the masses of the sterile neutrino and left handed neutrino comparable, an example of the hierarchy problem.  Given the contrived nature of organising the matter content in the SM, it seems natural  for us to consider the role of heavy right-handed neutrinos in leptogenesis.
 

 \subsection{Dark matter and the axion}
 
  The amount of visible baryonic matter in the universe is not enough to fit observations:
  \begin{itemize}
  \item The initially overlooked paper by Zwicky~\cite{Zwicky} on the velocity dispersion of galaxies in the Coma cluster led him to postulate additional Dark matter.
      \item On cosmological scales the density of the universe is at the critical density according to observations.The component of visible matter is only $4\%$ of the critical density. Other contributions are needed; the dark matter contribution is estimated to be about $25\%$~\cite{Planck,BCluster, Profumo}.
  \item Dark matter in the early universe is also required for formation of structure in the universe~\cite{Primack, Structure}.
\end{itemize}
We will be interested in pseudoscalar light particles $b(x)$ which interact with generic gauge fields through the field strength $G_{\mu \nu }$ (a matrix in some internal symmetry space).  There could be more than one $G_{\mu \nu }$ associated with different gauge fields. The coupling of the pseudoscalar to the gauge fields is a specific term $L_{b}$ in the Lagrangian:
 \begin{equation}
\label{ }
L_{b}\sim b\left( x\right)  Tr\left( G^{\mu \nu }\tilde{G}_{\mu \nu } \right)  
\end{equation}
where $\tilde{G}_{\mu \nu } \equiv \epsilon_{\mu \nu \gamma \delta } G^{\gamma \delta }$. We will call such a pseudoscalar an axion. Such axions can have a role as dark matter~\cite{Marsh} and are associated with high energy physics and the strong CP problem~\cite{Peccei}; but here we will consider an axion arising from gravitational physics associated with String theory. 

 \subsection*{Outline}
We shall discuss the antilinear $\mathcal{CPT}$ symmetry  within a string theory context. It will be argued that the breaking of Lorentz symmetry by a homogeneous background related to the Kalb-Ramond~\cite{Kalb, R16} axion field will lead to  $\mathcal{CPT}$ violation (CPTV). This CPTV background will be essential for producing leptogenesis in our model. Moreover quantum  fluctuations of the axion will lead to the Kalb-Ramond axion being coupled to the axial anomaly expressed in terms of gauge fields in the model. We shall comment on the role of this axion as dark matter.

 \section{ The model}

In seeking physics beyond the SM to solve the baryogengesis problem, we will consider first leptogenesis~\cite{R2}. This is attractive since one scenario for leptogenesis uses the seesaw mechanism for neutrino mass generation. Sphaleron processes~\cite{Gorbunov} then convert lepton number asymmetry into baryogenesis~\cite{R13}. In such a scenario lepton asymmetry is generated first by means of decays of right handed sterile Majorana neutrinos to SM particles which freeze out at a certain temperature because of the expanding universe. This scenario has been shown to be viable\cite{R10Boss,R11Boss,R12Mav,R13Mav}.
In general terms,  gravity at Planck energies cannot be ignored since the gravitational coupling constant $G_{N}$ is of order $M_{P}^{-2}$ where $M_{P}$ is the Planck mass. At an energy scale $E$ the dimensionless constant  is $G_{N} E^{2}$ and so, at least at the Planck scale, quantum gravity effects cannot be ignored. Within a string theory framework the gravitational sector\footnote{The lowest energy states in a superstring theory are the graviton, dilation and the Kale-Ramond field~\cite{R17}.} also has the Kalb-Ramond field and the presence of a Kalb-Ramond background would break CPT invariance. 
From experiments there are stringent upper bounds~\cite{R3, R4} in the present era for CPTV. However CPTV could have been much greater in the early epoch of the universe, a possibility with consequences that we will explore. In our model we can show how the background diminishes with temperature (and hence time) so as to easily satisfy experimental constraints.


\subsection{String theory background}

  Within bosonic string theory, in the closed string sector, the spectrum contains three massless states $g_{\mu\nu}$, the graviton, $B_{\mu\nu}$, the Kalb-Ramond (KR) antisymmetric field and $\phi$ the dilaton\footnote{In superstring theory any tachyonic states present in the bosonic formulation are eliminated.}. In  an effective field theoretic approximation to strings, the  low energy excitations of strings, have their own fields. Strings do not have to be formulated in flat backgrounds and in particular they can be formulated on backgrounds formed from gravitons, KR fields and dilatons.  For consistency of the world sheet formulations of string theory it is necessary to have conformal invariance~\cite{Polchinski1, Polchinski2}. This leads to the vanishing of beta functions calculated through the renormalisation group. Effective actions can be constructed for these background fields so that their equations of motion are equivalent to the vanishing of beta functions. The calculations of these beta functions can be done in a Regge slope ($\alpha^{\prime } $ ) expansion.
  
  The requirement of conformal invariance leads to target space dimensions higher than four. Dimensional reduction is employed~\cite{Witten85, Ibanez, Gasperini} in order to make contact with the currently observed universe.~On spacetime reduction to four dimensions, the bosonic part of the (four space-time dimensional) effective action, $S_B$, reads in the Einstein frame~\cite{R18, R19, R20}:
\begin{align}\label{sea2}
S_B =&\; \dfrac{1}{2\kappa^{2}}\int d^{4}x\sqrt{-g}\Big(R - e^{-4\Phi}H_{\lambda\mu\nu}H^{\lambda\mu\nu} - \Omega\Big) + \dots,
\end{align}
where
\begin{equation}\label{hfield}
H_{\mu\nu\rho} = \partial_{[\mu}\, B_{\nu\rho]},
\end{equation}
and  $\kappa^2 = 8\pi G_{N}$, and $G_{N}={ M_P}^{-2}$ is the (3+1)-dimensional Newton constant (with $M_{P}= 1.22 \times 10^{19}$~GeV the four-dimensional Planck mass). The $\dots$ denote in particular higher derivative terms in the dilaton field and $\Omega$ is related to the cosmological constant. The constant $G_{N}$ is related to the string mass scale $M_s$ via~\cite{R17}: ${G_{N}}^{-1} = {\mathcal V}^{(n)} \,  M_s^{2+n}$, with ${\mathcal V}^{(n)}$ a compactification volume (or appropriate bulk volume factor, in brane-universe scenarios). 
In the closed string sector, $B_{\mu\nu}$ is a gauge field with a  gauge symmetry 
$B_{\mu\nu} \rightarrow B_{\mu\nu} + \partial_\mu \theta_\nu - \partial_\nu \theta_\mu$ which characterises the target-space effective action. This implies that
the action depends only on the field strength of the field $B_{\mu\nu}$, which is a three-form with components
\begin{equation}\label{hfield}
H_{\mu\nu\rho} = \partial_{[\mu}\, B_{\nu\rho]},
\end{equation}
where the symbol $[\dots ]$ denotes complete antisymmetrisation of the respective indices. 
The 3-form $H_{\mu\nu\rho}$ satisfies classically the Bianchi identity 
\begin{equation}\label{bianchi}
\partial_{[\mu}\, H_{\nu\rho\sigma]} = 0, 
\end{equation}
by construction.~\footnote{In (Heterotic) string theory, in the presence of gauge and gravitational fields, the right-hand-side of (\ref{hfield}) is modified by appropriate (parity-violating) Chern--Simons three-forms, which lead to a non-zero right-hand side of the Bianchi identity (\ref{bianchi}), expressing gauge and gravitational anomalies~\cite{R17}. }

In our model torsion will also be considered at a quantum level and the chiral anomaly (which will have gravitational and gauge contributions through topological terms) will lead to a direct coupling of the axion to these topological densities. 
Before considering mechanisms for producing $CPT$-violating backgrounds,  
we will introduce and consider our phenomenological model.
\section{Leptogenesis model}
 In \cite{R9, R10} we  considered leptogenesis originating from the \emph{tree-level} decay of a heavy sterile (right-handed, Majorana) neutrino (RHN)  into SM leptons, in the presence of generic CPTV time-like axial backgrounds, assumed constant in the cosmological (Robertson-Walker) frame of the early universe. The resulting Lagrangian is given by: 
\begin{equation}
\label{e2}
{\mathcal{L}}= {\mathcal L}_{\rm SM} + i\overline{N}\slashed{\partial}N-\frac{m_{N}}{2}(\overline{N^{c}}N+\overline{N}N^{c})-\overline{N}\slashed{B}\gamma^{5}N-\sum_k \, y_{k}\overline{L}_{k}\tilde{\varphi}N+h.c.
\end{equation}
where ${\mathcal L}_{\rm SM}$ denotes the SM Lagrangian, $B_\mu$ is a CPTV background c-number field, associated with physics beyond the SM, 
$N$ is the RHN field with (Majorana) mass $m_N$,  $\tilde \varphi$ is the adjoint ($\tilde{\varphi}_i=\varepsilon_{ij}\varphi_j $) of the Higgs field  $\varphi$, 
 and $L_{k}$ is a lepton (doublet) field of the SM sector, with $k$ a generation index. $y_k$ is a Yukawa coupling, which is non-zero and provides a non-trivial (``Higgs portal'') interaction between the RHN and the SM sectors. In the case of \cite{R9, R10} a single sterile neutrino species 
suffices to generate phenomenologically relevant lepton asymmetry, and hence from now on 
we restrict ourselves to the first generation ($k=1$), and set 
\begin{equation}\label{e3}
y_1 = y ~.
\end{equation}
In the scenario of \cite{R9, R10}, the CPTV background $B_\mu$ is assumed to have only a non-zero temporal component~\cite{R21}, which was taken to be constant in the Robertson-Walker frame of the early universe, 
\begin{equation}\label{temporalB}
B_0 = {\rm const} \ne 0~, \, B_i = 0 ~, i=1,2,3~.
\end{equation}
In this case, the Lagrangian (\ref{e2}) assumes the form of a SME Lagrangian in a Lorentz and CPTV background~\cite{R15}. 

A lepton asymmetry is then generated due to the CP and CPTV tree-level decays of the RHN $N$ into SM leptons, 
in the presence of the background (\ref{temporalB}), induced by the Higgs portal Yukawa interactions of (\ref{e2})~\cite{R9,R10}:
\begin{eqnarray}\label{4channels}
{\rm Channel ~I}&:& \qquad  N \rightarrow l^{-}h^{+}~, ~ \nu \, h^{0}~,  \\ \nonumber 
{\rm Channel ~II}&:& \qquad  N \rightarrow l^{+}h^{-}~,~  \overline \nu \, h^{0}~.
\end{eqnarray}
where $\ell^\pm$ are charged leptons, $\nu$ ($\overline \nu$) are light, ``active'', neutrinos (antineutrinos) in the SM sector,
$h^0$ is the neutral Higgs field, and 
 $h^\pm$ are the charged Higgs fields\footnote{At high temperatures, above the spontaneous electroweak symmetry breaking, the charged Higgs fields $h^\pm$ do not decouple from the physical spectrum, and play an important r\^ole in leptogenesis.}.  As a result of the non-trivial $B_0 \ne 0$ background (\ref{temporalB}), the decay rates of the Maorana RHN between the channels I and II are different, resulting in a Lepton asymmetry, $\Delta L^{TOT}$, which then freezes out a temperature $T_D$. In \cite{R10, R11}, a detailed study of the associated Boltzmann equations for the processes in (\ref{4channels}), and their reciprocals,  has led to the result:
\begin{equation}\label{totDL}
\dfrac{\Delta L^{TOT}}{s} \simeq  (0.016, \, 0.019) \,  \dfrac{B_{0}}{m_{N}},  
\end{equation}
$ {\rm at~the~ freezeout~temperature} \quad T=T_D : \quad m_N/T_D  \simeq (1.44, \, 1.77),$where $s$ is the entropy density of the universe (and the numbers inside the  brackets distinguish different Pad$\acute{e}$ approximations). This implies the phenomenologically acceptable values of the lepton asymmetry of ${\mathcal O}(8 \times 10^{-11})$, which can then be communicated to the baryon sector via Baryon-minus-Lepton-number ($B-L$) conserving sphaleron processes~\cite{R13} in the SM, thus producing the observed amount of baryogenesis  in the Universe,  occur for values of
\begin{equation}\label{b0}
\frac{B_0}{m_N} \sim  10^{-9}, \quad {\rm at~ freezeout~temperature} \quad T=T_D : \quad m_N/T_D  \simeq (1.77, 1.44),
\end{equation}
The different values $(a,b)$ of the numerical coefficients in the right-hand-side of  the two equations in (\ref{totDL}), are due to two different analytical methods (series expansion ($a$) and integrating factor ($b$) method~\cite{R10, R11}, respectively) used in the Pad$\acute{e}$ approximant solution of the Boltzmann equations associated with (\ref{4channels}). With the value $y \sim 10^{-5}$ of the Yukawa coupling (\ref{e3}) , and for $m_N = {\mathcal O}(100)$~TeV~\cite{R9, R10} we  obtain a $B_0 \sim 0.1~{\rm MeV}$, for phenomenologically relevant leptogenesis to occur at $T_D \simeq (56 - 69) $ TeV, in our scenario.
In \cite{R9, R10} the microscopic origin of the background $B_0$ was not discussed in detail.

\section{Towards a microscopic understanding of the background}
In (\ref{sea2}) the  derivatives of the dilaton field, $\Phi$, will be ignored since the dilaton is assumed to be slowly varying~\cite{R9, R10}  
at late epochs of the universe, relevant for leptogenesis; hence we may  approximate $\Phi \simeq {\rm constant}$, which can thus be absorbed in appropriate normalisations of the KR field. In this approximation, 
the vacuum energy term $\Omega$ is treated as a constant, to be determined phenomenologically by requiring appropriately suppressed vacuum energy contributions. 

It is known~\cite{R17, R18, R19} that the classical KR field strength terms $H^2$ in (\ref{sea2}) can be absorbed into a generalised curvature  with a ``torsionful connection (Christoffel symbol)'', with the contorsion proportional to the field strength $H_{\mu\nu}^\rho$ , 
${\overline \Gamma}_{\mu\nu}^{\rho} = \Gamma_{\mu\nu}^\rho +  H_{\mu\nu}^\rho  \ne {\overline \Gamma}_{\nu\mu}^{\rho}$,
where $\Gamma_{\mu\nu}^\rho = \Gamma_{\nu\mu}^\rho$ is the torsion-free Christoffel symbol. In the string effective action this is only an approximation since there are higher order corrections in $\alpha^{\prime } $. Nonetheless it is a useful starting point.This is our  motivation to consider fermions on a torsional manifolds in a microscopic model for our background~\cite{R10,R11,R23,R24,R25}. Lorentz invariance does not hold in the presence of a torsion background.

\subsection{Fermions and torsion}

We will make a connection with the Einstein-Cartan theory\cite{R5} (EC) (see appendix \ref{EC}) which will model the above approximation to our string theory effective actions. Torsion in EC is a non-propagating field and the gravitational dynamics is described comparatively simply through a generalised scalar curvature term (unlike in string effective actions which have higher derivative terms). We will consider the fermions in the SM moving in this torsion-gravitational background. The gravitational part of the action is 
\begin{equation}
\label{E1}
S_{\rm EC}  = \frac{1}{8\pi G} \int d^{4}x\sqrt{g}  \bar{R} \left( \varpi \right)   
\end{equation}
where $\bar{R}$ is the generalised scalar curvature in terms of the torsion modified Christoffel symbol.
 We will first  write down the action~\cite{R10,R11,R25}  $ S_{\psi }$ for an electrically charged Dirac fermion in this torsionful  background~\cite{R6,R7,R8}:
\begin{eqnarray}
S_{\psi} & = & \int d^{4}x\sqrt{-g}\left[\frac{i}{2}\left(\overline{\psi}\gamma^{\mu}\overline{D}\left(\overline{\omega}\right)_{\mu}\psi-\left(\overline{D}\left(\overline{\omega}\right)_{\mu}\psi\right)\gamma^{\mu}\psi\right)-m\overline{\psi}\psi\right] \label{E2}\\
 & = & \int d^{4}x\sqrt{-g}\overline{\psi}\left(i\gamma^{\mu}\partial_{\mu}-m\right)\psi+\int d^{4}x\sqrt{-g}\left(F_{\mu}+B_{\mu}\right)\overline{\psi}\gamma^{5}\gamma^{\mu}\psi \label{E3}
\end{eqnarray}
\begin{equation}
\label{E4}
\overline{D}_{a}=\partial_{a}-\frac{i}{4}\overline{\omega}_{bca}\sigma^{bc}, \qquad \sigma^{ab}=\frac{i}{2}\left[\gamma^{a},\gamma^{b}\right],
\end{equation}
\begin{equation}
\label{E5}
F^{\mu}=\varepsilon^{abc\mu}e_{b\lambda}\partial_{a}e_{c}^{\lambda}, \qquad B^{\mu}=-\frac{1}{4}e^{-2\phi}\varepsilon_{abc}^{\quad\;\mu}H^{abc} ,\qquad J^{5\mu}=\overline{\psi}\gamma^{\mu}\gamma^{5}\psi.
\end{equation}
$\eta_{ab}$ is the flat space metric on the tangent plane at  the spacetime point $x$,$\varGamma_{\mu\sigma}^{\nu}$ is the Christoffel symbol and $e_{\nu}^{a}$ is the vierbein\footnote{Latin indices will denote tensors in the tangent space basis and Greek indices indices in the space-time manifold. A vierbein relates a quantity with Latin indices to one with Greek indices,~e.g. $H^{abc}=\eta^{a}_{\mu } \eta^{b}_{\nu } \eta^{c}_{\lambda } H^{\mu \nu \lambda }$ .}. The interaction of the sterile neutrino with the background field \eqref{e2} is similar to that found in this microscopic derivation. For the Friedmann-Robertson-Walker (FRW) metric\cite{Dodelson} $F^{\mu}$ vanishes. The vierbeins determine the metric through $g_{\mu\nu}=\eta_{ab}e_{\mu}^{a}e_{\nu}^{b}$. In the absence of torsion the spin connection $\omega_{\mu}^{\;ab}$ is
\begin{equation}
\label{E6}
\omega_{\mu}^{\;ab}\equiv e_{\nu}^{a}\left[\partial_{\mu}e^{\nu b}+\varGamma_{\mu\sigma}^{\nu}e^{\sigma b}\right].
\end{equation}
This is modified in the presence of torsion to 
\begin{equation}
\label{E7}
\varpi_{ab\mu } =\omega_{ab\mu } +K_{ab\mu }
\end{equation}
where $K_{abc}=\frac{1}{2} \left( H_{cab}-H_{abc}-H_{bca}\right) $ is the contortion.
In \emph{four space-time} dimensions the KR three-form $H$ can be written as
\begin{equation}
\label{E8}
\partial^{\mu } b=-\frac{1}{4} \exp \left( -2\varphi \right)  \epsilon^{\  \  \  \  \  \mu }_{abc} H^{abc}.
\end{equation}
$b$ will be identified with the axion field and is a feature of a 4-dimensional spacetime. Our symmetry group  can be generalised to include the $SU(3)$ colour gauge symmetry, e.g. $SU\left( 3\right)_{c}  \times U\left( 1\right) $ . All fermion species will contribute to the axial current. This generalisation is necessary since it is pertinent to the role of the KR field in axion cosmology and the  strong CP problem.

\subsection{Quantum torsion in Einstein-Cartan theory}
In order to avoid a proliferation of indices it is convenient to use the language of differential forms. We have two independent 1-forms:

\begin{equation}
\label{E9}
{\bf{e}}^{a}\equiv e^{a}_{\mu }\left( x\right)  dx^{\mu },\  \  \ {\bf{ \omega^{a}_{b}}} \equiv \omega^{a}_{b\mu } \left( x\right)  dx^{\mu }.
\end{equation}
There are two associated 2-forms.  The 2-forms ${\bf{T}}^{a}$ for torsion and ${ \bf{R}}^{a}_{b}$ for curvature are given by\footnote{A $p$-form ${\bf{A}}$ has the representation\[{\bf{A}}=A_{\left[\mu_{1}\ldots\mu_{p}\right]}dx^{\mu_{1}}\wedge\ldots dx^{\mu_{p}}\] where the summation convention is in place. If ${\bf{A}}$ is a $p$-form and ${\bf{B}}$ is a $q$-form then we have the ($p+q$)-form ${\bf{C}}={\bf{A}}\wedge {\bf{B}}$ given by
\[
{\bf{C}}=A_{\mu_{1}\ldots\mu_{p}}B_{\mu_{p+1}\ldots\mu_{p+q}}dx^{\mu_{1}}\wedge\ldots dx^{\mu_{p+q}}.
\]
In $D$-dimensions a Hodge dual $^{*}{\bf{A}}$ of ${\bf{A}}$ is a $(D-p)$-form
\[
^{\ast}{\bf{A}}=\frac{1}{(D-p)!}A^{\mu_{1}\ldots\mu_{p}}\eta_{\mu_{1}\ldots\mu_{p}\mu_{p+1}\ldots\mu_{D}}dx^{\mu_{P+1}}\wedge\ldots dx^{\mu_{D}}
\]
where $\eta_{\mu_{1}\ldots \mu_{D}}$ is a weight $0$ tensor.}:

\begin{equation}
\label{E10}
{\bf{T}}^{a}\equiv de^{a}+\omega^{a}_{b} \wedge e^{b},\  \  \ { \bf{R}}^{a}_{b}=d\omega^{a}_{b} +\omega^{a}_{c} \wedge \omega^{c}_{b} .
\end{equation}
The contorsion 1-form $K^{ab}$ is by definition 
\begin{equation}
\label{E11}
K^{ab}=\omega^{ab}-\tilde \omega^{ab}
\end{equation}
where $\tilde \omega^{ab}$ is the spin connection in the absence of torsion. Hence, by definition, $T^{a}=K^{a}_{b}\wedge e^{b}.$
The Einstein-Cartan gravitational part of the action is
\begin{equation}
\label{E12}
S_{EC}=\frac{1}{4\kappa^{2}}\int\varepsilon_{abcd}\mathbf{\mathcal{R}}^{ab}\wedge\mathbf{e}^{c}\wedge\mathbf{e}^{d}
\end{equation}
The torsion tensor can be decomposed into irreducible representations as follows:
\begin{equation}
\label{E13}
T_{\mu \nu \rho }=\frac{1}{3} \left( T_{{}\nu }g_{\mu \rho }-T_{\rho }g_{\mu \nu }\right)  -\frac{1}{3!} \epsilon_{\mu \nu \rho \sigma } S^{\sigma }+q_{\mu \nu \rho }
\end{equation}
with $\epsilon_{\mu \nu \rho \sigma } q^{\nu \rho \sigma }=q^{\nu }_{\rho \nu }=0$\footnote{We are free to switch between Greek and Latin indices as we choose.}. It is $S^{\sigma }$ which couples to the axial fermion current through the following term in the action~\cite{R9}:
\begin{equation}
\label{E14}
-\frac{3}{4} \int {\bf{S}}\wedge {\bf{^{*}J}}^{5}\in S_{\psi }.
\end{equation}
We have seen this coupling to the axial current already in \eqref{E3}.
The contorsion tensor has the following decomposition:
\begin{equation}
\label{E15}
K_{abc}=\frac{1}{2} \epsilon_{abcd} S^{d}+\hat{K}_{abc} .
\end{equation}
$\hat{K}$ includes the trace vector $T_{\mu}$ and the tensor $q_{\mu\nu\rho}$ parts of the torsion tensor.~$S_{EC}$ can be rewritten in terms of the Einstein Ricci scalar as follows\begin{equation}
\label{E16}
S_{G}=\frac{1}{2\kappa^{2} } \int d^{4}x\sqrt{-g} \left( R+\hat{\Delta } \right)  +\frac{3}{4\kappa^{2} } \int {\bf{S}}\wedge {\bf{S^{\ast}} }
\end{equation}
where $\hat{\Delta } =\hat{K}^{\lambda }_{\  \  \mu \nu } \hat{K}^{\nu \mu }_{\  \  \  \lambda } -\hat{K}^{\mu \nu }_{\  \  \  \nu } \hat{K}^{\  \  \  \lambda }_{\mu \lambda } .$ So far we have rewritten the Einstein-Cartan action and indicated how fermions can couple to gravity including torsion~\cite{R9}. In String theory the KR excitation could be expected to have quantum fluctuations since the excitation appears in the spectrum of the quantum string. Within our framework we will quantise the torsion (but keep the gravitational field classical) by using path integrals.
 From the classical equation of motion we find
\begin{equation}
\label{E17}
K_{\mu ab}=-\frac{1}{4} e^{c}_{\mu }\epsilon_{abcd} \bar{\psi } \gamma_{5} \gamma^{d} \psi 
\end{equation}
which implies $d^{\ast}{\bf{S}}=0$ and that $Q=\int {\ast}S$ is conserved. Following Duncan et al \cite{R25} we will postulate that this geometric conservation is maintained at the \emph{quantum} level by adding a factor $\delta \left( d^{\ast }S\right) $ in the path integral using a Lagrange multiplier\footnote{ Maintaining this constraint in the quantum theory may require counterterms. }.The \emph{torsion part} of the quantum gravity path integral is
\begin{eqnarray}
Z & = & \int DSDb\exp \left[ i\int \frac{3}{4\kappa^{2} } S\wedge \ast S-\frac{3}{4} S\wedge \ast J^{5}+\left( \frac{3}{2\kappa^{2} } \right)^{1/2}  bd\ast S\right] \label{E18}\\
 & = & \int Db\exp \left[- i\int \frac{1}{2} db\wedge \ast db+\frac{1}{f_{b}} db\wedge \ast J^{5}+\frac{1}{2f^{2}_{b}} J^{5}\wedge \ast J^{5}\right] \label{E19}  
\end{eqnarray} 
where $f_{b}=\sqrt{\frac{8}{3\kappa^{2} } } $. We may partially integrate the second term in the exponential in \eqref{E18} and note that the axial current has anomalies~\cite{R26, R27, R12}. 

Let us return to the full path integral which we write as 
\begin{equation}
\label{ee18}
Z=\int \mathcal{D}g\mathcal{D}\psi \mathcal{D}\bar{\psi } \mathcal{D}b\exp \left[ iS_{eff}\right]  .
\end{equation}
In realistic situations 
\begin{equation}
\label{Ee1}
{}_{}J^{5}_{\mu }=\sum^{f}_{i=1} \bar{\psi }_{i} \gamma_{\mu } \gamma^{5} \psi_{i} .
\end{equation}
 We will split the $b$ field into a quantum  $\tilde b$ and classical part $\bar b$; the effective action (in conventional notation) can be written as 
\begin{eqnarray}
S_{eff} & = & \frac{1}{2\kappa^{2} } \int d^{4}x\sqrt{-g} \left( R+\frac{8}{3} \partial_{\sigma } \bar{b} \partial^{\sigma } \  \bar{b} -\Omega \right)  +S_{free}-\int d^{4}x\sqrt{-g} \partial_{\mu } \bar{b} J^{5\mu } \nonumber\\
&  & -\frac{3\kappa^{2} }{16} \int d^{4}x\sqrt{-g} J^{5}_{\mu }J^{5\mu }+\frac{8}{3\kappa^{2} } \int d^{4}x\sqrt{-g} \partial_{\sigma } \bar{b} \  \partial^{{}^{\sigma }} \tilde{b} 
 \nonumber\\
 &  & +\frac{1}{2\kappa^{2} } \int d^{4}x\sqrt{-g} \frac{8}{3} \partial_{\sigma } \tilde{b} \partial^{\sigma } \tilde{b} +\frac{1}{\kappa }\int d^{4}x\sqrt{-g} \tilde{b} \partial_{{}\mu } J^{5\mu }\label{fluctuations}
\end{eqnarray}
 where $S_{free}$ is the action  of a free fermion in  a gravitational background.
  \section{Consequences}
 Equation \eqref{fluctuations} is the master equation of our theory.
 Let us examine the consequences of the quantum fluctuations of the axion that we can deduce from \eqref{fluctuations}.
 
 \begin{itemize}
  \item Consider an axion background $\bar b(x)$ in the FRW background which is linear in (cosmic) time and satisfies the conformal invariance conditions in non-critical strings to be a background (see \cite{R21}). This background decouples from $\tilde b$ since the $\bar b \tilde b$ term vanishes (provided $\int d^{4}x\  \partial^{0} \bar{b}$ vanishes).The term $\int d^{4}x\sqrt{-g} \partial_{\mu } \bar{b} J^{5\mu }$ gives the model of leptogenesis represented by \eqref{e2} with a constant axion background. We should sound a word of caution since the constant axion background may not be solution of superstring theory (in the presence of fermions). Even if it were a solution, the constant would need to be fine tuned.
  
 This case has another use , besides leptogenesis, which is for axionic dark matter and a solution of the strong CP problem~\cite{Peccei}.
 In terms of torsionless quantities the axial current anomaly is
\begin{equation}
\label{E20}
d{\bf{^{\ast }J^{5}}}=-\frac{\alpha_{QED} Q^{2}}{\pi } {\bf{F}}\wedge {\bf{F}}-\frac{\alpha_{s} N_{q}}{2\pi } Tr\left( {\bf{G}}\wedge {\bf{G}}\right)  -\frac{N_{f}}{8\pi^{2} } {\bf{R}}^{ab}\wedge {\bf{R}}_{ab}
\end{equation}
where $\alpha_{QED}$ and $\alpha_{s}$ are the fine structure constants for electromagnetism and quantum chromodynamics respectively, $N_f$ is the number of fermion flavours and $Q^{2}=\sum_{f} Q^{2}_{f}$ where $Q_f$ is the electric charge of the fermion with flavour $f$, $F$, $G$ and $R$ are the field strength forms of QED, QCD and gravity. The resulting effective action  is:
\begin{eqnarray}
\label{E21}
S_{eff}&=&S_{0}-\frac{1}{2f^{2}_{b}} \int J^{5}\wedge \ast J^{5}-\frac{\alpha_{QED} }{\pi f_{b}} \int bF\wedge F\\
& &-\int \frac{1}{2} db\wedge \ast db-\frac{1}{8\pi^{2} } \int \left( \Theta +\frac{N_{f}}{f_{b}} b\right)  R^{ac}\wedge R_{ac}\nonumber\\
&& -\frac{\alpha_{s} }{2\pi } \int \left( \theta+\frac{N_{q}}{f_{b}} b\right)  Tr\left[ G\wedge G\right] \nonumber 
\end{eqnarray}
The constants $\Theta$ and $\theta$ are associated with topological terms and have been kept in for completeness (and classically do not affect the equations of motion).~$S_{0}$ represents the Einstein gravitational interactions, non-gravitational  gauge interactions and Yukawa interactions of matter. This model embodies the links that we mentioned in the introduction, and the $b$ field is at the heart of the connections.
Using the action in (31) we have explored leptogenesis. Recently the formulation used in \eqref{E21} has also been investigated in connection with  axionic dark matter. Although the model has features similar to that of the QCD axion, the gravitational axion couples to the Pontryagin densities of all the gauge interactions in the model. The model could, for example, be a generalisation of the Standard Model above the electroweak transition; then all the gauge fields would have couplings to the axion through their Pontryagin density. Another feature is the repulsive current -current interaction. This would be a contribution to the de Sitter nature of the Universe which is however suppressed by the factor of $f_b^2$ where $f_b=\frac{M_p}{\sqrt{3\pi}}\approx 4\times 10^{18}GeV$ (and $M_p$ is the Planck mass). It has been estimated~\cite{R24} that this gravitational axion would have a mass  of order $10^{-12 }eV$. Axions can have a range of masses and such a mass has not been ruled out.

  \item In the scenario for the Kalb-Ramond torsion model that we have just discussed,  we consider backgrounds of $B_{0}$ ($=\partial_{0}b$) which are constant but then typically we  argue that  the $B_{0}$ has diminished from the time of freeze-out of the sterile neutrino in our model to values which are compatible with current bounds for CPTV. It has more recently been considered that $B_{0}$ can be time or temperature $T$ dependent.\footnote{In the radiation era the scale factor $a(t)$ in the Friedmann-Robertson-Walker metric goes with time like $a(t)\sim t^{1/2}$. }
  From (31)  we find the equation of motion of $b$ to be
 \begin{equation}
\label{E22}
\partial_{\mu } \left[ \sqrt{-g} \left( \frac{8}{3\kappa^{2} } \partial_{\mu } b-J^{5\mu }\right)  \right]  =0.
\end{equation}
For the flat FRW metric   $\sqrt{-g} \sim a^{3}\sim T^{-3}$ where $g$ is the determinant of this metric and $a(t)$ is the scale factor. We consider for \eqref{E22} only dependence on time.
On assuming that a chiral current condensate has not formed we find that 
\begin{equation}
\label{E23}
B_{0}\left( T\right)  \sim T^{3}.
\end{equation}
If we assume that the decoupling temperature $T_{D}=O\left( 100\right) $ TeV, as was the case for constant $B_{0}$ analysis, we are in the radiation era. In the general case 
\begin{equation}
\label{ }
\frac{8}{3\kappa^{2} } B_{o}\left( T\right)  -\langle J^{5}_{0}\rangle_{T} =A^{\prime }T^{3}
\end{equation}
where $A'$ is a constant of integration. We can evaluate the thermal average $\langle J^{5}_{0}\rangle_{T}$ using equilibrium thermodynamics in the presence of a background field $B_{0}(T)$. Let us write
 \begin{equation}
\label{T1}
J^{5}_{0}=\left< J^{5}_{0}\right>_{T}  +\sum_{i=fermions} \psi^{\dag }_{i} \gamma^{5} \psi_{i}. 
\end{equation}
The fermion part of the Lagrangian  is 
\begin{equation}
\label{T2}
L^{F}=\sqrt{-g} \bar{\psi } \left[ i\gamma^{\mu } \partial_{\mu } -m1-\left( B_{0}\left( T\right)  -\mu_{5} \left( T\right)  \right)  \gamma^{0} \gamma^{{}^{5}} +\mu \left( T\right)  \gamma^{0} \right]  \psi +\ldots 
\end{equation}
with $\mu_{5} \left( T\right)  =\frac{3\kappa^{2} }{8} \left< J^{5}_{0}\right>_{T}  .$The  $\cdots$ denotes current-current interactions which are highly suppressed and will be ignored. A chemical potential $\mu$ has been added for the quarks. For leptons, since lepton number is not conserved, the chemical potential is $0$. We will consider the situation when there are no external electromagnetic fields. The dispersion relation for a fermion of mass $m$ and helicity $\lambda_{r}$ is 
\begin{equation}
\label{T3}
E_{r}=\sqrt{m^{2}+\left( \left| \vec{p} \right|  +\lambda_{{}_{r}} B_{0}\right)^{2}  } .
\end{equation}
The thermal distributions $f(E)$ depend on the chemical potential:
\begin{equation}
\label{T4}
f\left( E\right)  =\frac{1}{\exp \left( \frac{E-\mu }{T} \right)  +1} .
\end{equation}
For a  particle with (left/right) handedness $\mu =\mu_{L\left( R\right)  } =\mu -\left( +\right)  \mu_{5} $.The thermal distributions are used in the calculation of $\left< J^{5}_{0}\right>_{T}$. For high temperatures $T\ge T_{D}$, where $T_D$ is the decoupling temperature of the sterile neutrino, we can consider massless fermions and so $E_{r}=\left| \left| \vec{p} \right|  +\lambda_{r} B_{0}\right|  $. For antiparticles $E\to -E$ and $\{\mu, \mu_{5} \to -\mu, -\mu_{5} \}$. The thermal average $\left< J^{5}_{0}\right>_{T}$ has contributions from all species of fermions, both leptons and quarks.
\begin{eqnarray}
\left< J^{5}_{0}\right>_{T} & = & \sum_{i} \left< \bar{\psi }_{i} \gamma^{0} \gamma^{{}  5} \psi_{i} \right>_{T}   = {}^{}_{}\sum_{i,\lambda } \left[ \left( n^{eq}_{R}-\bar{n}^{eq}_{R} \right)  -\left( n^{eq}_{L}-\bar{n}^{eq}_{L} \right)\right]_{i,\lambda } \\
 & = &  \sum_{i}\{    \left[ \left( n^{eq}_{R}-\bar{n}^{eq}_{R} \right)  -\left( n^{eq}_{L}-\bar{n}^{eq}_{L} \right)\right]_{\lambda=-1 }+ \left[ \left( n^{eq}_{R}-\bar{n}^{eq}_{R} \right)  -\left( n^{eq}_{L}-\bar{n}^{eq}_{L} \right)\right]_{\lambda=1 }    \}\nonumber\\
     & = & \sum_{i}\{    \left[ \left( n^{eq}_{R}-\bar{n}^{eq}_{R} \right)  -\left( n^{eq}_{L}-\bar{n}^{eq}_{L} \right)\right]_{\lambda=-1 }+ \left[ \left( n^{eq}_{L}-\bar{n}^{eq}_{L} \right)  -\left( n^{eq}_{R}-\bar{n}^{eq}_{R} \right)\right]_{\lambda=-1 }\} \nonumber\\
 & = & 0 \nonumber
\end{eqnarray} where $n^{eq}  (\bar{n}^{eq})$  is the thermal equilibrium number density of particles and antiparticles and the sum is over all species of fermions as well as the helicities. Hence the thermal axial current expectation vanishes and 
\end{itemize}
 $$ B_{0}\left( T\right)  \simeq AT^{3},\  \  \  \  A=\frac{3\kappa^{2} }{8} A^{\prime }$$ 
 where $A^{\prime }$ is a constant of integration. $A$ is determined by the boundary value $B_{0}\left( T=T_{D}\sim 100TeV\right) $ which is chosen to give  phenomenologically acceptable values of leptogenesis. Moreover
\begin{equation}
\label{T6 }
M^{-1}_{Pl}\frac{\dot{\bar{b} } }{\bar{b} } =-\frac{3}{2} \frac{H\left( t\right)  }{M_{Pl}} \ll 1,\  \  \  \  \  T\left( t\right)  \simeq T_{D}.
\end{equation} Thus the approximation of taking $B_0$ constant in our earlier work is a reasonable approximation.

\subsection*{Conclusions}
We have shown how the Kalb-Ramond  field can be interpreted as a torsion with quantum fluctuations in a model for leptogenesis. This same model allows a possible resolution of the strong CP problem and provides a candidate for dark matter.~It would be interesting to see whether the phenomenology of leptogenesis, the strong CP problem and dark matter  are mutually compatible.

\section*{Acknowledgments}

The work of S.S. is supported in part by the UK Science and Technology Facilities research
Council (STFC) and UK Engineering and Physical Sciences Research
Council (EPSRC) under the research grants ST/T000759/1 and  EP/V002821/1, respectively. 

 \appendix{Einstein-Cartan theory}\label{EC}
 General relativity of Einstein is a geometric theory; the concept of inertial mass and gravitational mass coincide. In the SM gauge interactions there is no such equivalence for the particles' gauge charges. A theory very much like gravity can be obtained by gauging the Lorentz transformations of special relativity~\cite{R5, R6, R7, R8} by using vierbeins and spin (gauge) connections. 
Spacetime indices are labelled by Greek indices; Lorentz indices are labelled by Latin indices. The flat-spacetime (tangent space) Minkowski metric $\eta^{ab}$ is related to the manifold metric $g_{\mu \nu}$ by 
\begin{equation}
\label{e6}
g_{\mu\nu}=\eta_{ab}e_{\mu}^{a}e_{\nu}^{b}.
\end{equation}
We can make the decomposition $\omega^{ab}=\widetilde{\omega}^{ab}+\mathcal{K}^{ab}$ where $\mathcal{K}^{ab}$ is the contortion~\cite{R23} one form (and $\widetilde{\omega}^{ab}$ is the spin connection in the absence of torsion).

In terms of the contortion ${K}_{\:b}^{a}$ we also have
\[\mathcal{\mathcal{T}^{\mathit{a}}=K}_{\:b}^{a}\wedge e^{b}. \]
Let us define
\begin{equation}
\label{e7}
\widetilde{R}_{\:c}^{a}=d\widetilde{\omega}_{\:c}^{a}+\widetilde{\omega}_{\:b}^{a}\wedge\widetilde{\omega}_{\:c}^{b}\equiv\widetilde{D}\widetilde{\omega}_{\:c}^{a}.
\end{equation}
Hence 
    \[\mathcal{R}_{\:c}^{a}=\mathcal{\widetilde{R}}_{\:c}^{a}+\mathcal{\widetilde{D}}\mathcal{K}_{\;c}^{a}+\mathcal{K}_{\:b}^{a}\wedge\mathcal{K}^{bc}.\]
    In differential-form language the gravity action can be written as
    \begin{equation}
\label{e8}
S_{EC}=\frac{1}{4\kappa^{2}}\int\varepsilon_{abcd}\mathbf{\mathcal{R}}^{ab}\wedge\mathbf{e}^{a}\wedge\mathbf{e}^{b}
\end{equation}
 where    $\kappa^{2}=8\pi M_{P}^{2}$. For any fermionic field $\Psi$ we will take its  gauge covariant derivative to be 
\begin{equation}
\label{e9}
D\varPsi=d\varPsi+\frac{1}{4}\mathbf{\omega}^{ab}\gamma_{ab}\varPsi+ie\mathbf{A}\Psi+ig\mathbf{B}\Psi.
\end{equation}
Here we have just considered $\mathbf{A}$, the electromagnetic gauge 1-form, $\mathbf{B}$, the colour $SU(3)_{c}$ gauge 1-form and the Dirac gamma matrices $\gamma_{ab}=\frac{1}{2}\left[\gamma_{a},\gamma_{b}\right]$. (We will be mainly interested in the gravitational and colour degrees of freedom in relation to identifying the gravitational axion as a 
candidate for dark matter and as the  axion for the strong CP problem.) The fermionic action (on summing over flavours $f$) now reads
\begin{equation}
\label{e10}
\frac{1}{2}\sum_{f}\int\left(\overline{\varPsi}_{f}{\bf{\gamma}}\wedge*D\varPsi_{f}-D\overline{\varPsi}_{f}\wedge*{\bf{\gamma}}\varPsi_{f}\right)
\end{equation}
with $\mathbf{\gamma}=\gamma_{a}\mathbf{e}^{a}$ and $\overline{\Psi}=-i\Psi^{\dagger}\gamma^{0}$. The gauge kinetic energy contributions to the action are
\begin{equation}
\label{e11}
-\frac{1}{2}\int F\wedge*F-\int Tr\left[G\wedge*G\right]
\end{equation}
where $G$ and $F$ are the field strength 2-forms for the $SU(3)_{c}$ and $U(1)$ gauge fields.

\end{document}